\documentclass[12pt,preprint]{aastex}
\usepackage{latexsym,graphicx,natbib}

\def\simless{{\th \rlap{\raise 0.5ex\hbox{$\scriptstyle  {<}$}}
    {\lower 0.3ex\hbox{$\scriptstyle  {\sim}$}} \th }}  
\def\simgreat{{\th \rlap{\raise 0.5ex\hbox{$\scriptstyle  {>}$}}
    {\lower 0.3ex\hbox{$\scriptstyle  {\sim}$}} \th }}  
\def\greateq{{\th \rlap{\raise 0.5ex\hbox{$\scriptstyle  {>}$}}
    {\lower 0.3ex\hbox{$\scriptstyle  {-}$}} \th }}  
\def\lesseq{{\th \rlap{\raise 0.5ex\hbox{$\scriptstyle  {<}$}}
    {\lower 0.3ex\hbox{$\scriptstyle  {-}$}} \th }}  
\def\th{\thinspace}
\def\ts{{\raise 0.3ex\hbox{$\scriptstyle {\th \sim \th }$}}}

\usepackage{epsfig}
\newcommand{\etal}{\mbox{et al.}}
\newcommand{\saxjfull}{\mbox{SAX J1748.9$-$2021}}

\begin{document}

\shortauthors{Patruno \etal}
\shorttitle{Timing of \saxjfull\ }

\title{Phase-coherent timing of the accreting millisecond pulsar SAX J1748.9-2021}
\author{Alessandro Patruno \altaffilmark{1}, Diego Altamirano \altaffilmark{1}, Jason W.T. Hessels \altaffilmark{1},
Piergiorgio Casella \altaffilmark{1}, Rudy Wijnands \altaffilmark{1}, Michiel van der Klis \altaffilmark{1} }
\altaffiltext{1}{Astronomical Institute ``Anton Pannekoek,'' University of
Amsterdam, Kruislaan 403, 1098 SJ Amsterdam, The Netherlands, E-mail: apatruno@science.uva.nl}

\email{apatruno@science.uva.nl}

\begin{abstract}

We present a phase-coherent timing analysis of the intermittent
accreting millisecond pulsar \saxjfull. A new timing solution for the
pulsar spin period and the Keplerian binary orbital parameters was
achieved by phase connecting all episodes of intermittent pulsations
visible during the 2001 outburst. We investigate the pulse profile
shapes, their energy dependence and the possible influence of Type I
X-ray bursts on the time of arrival and fractional amplitude of the
pulsations. We find that the timing solution of \saxjfull\ shows an
erratic behavior when selecting different subsets of data, that is
related to substantial timing noise in the timing post-fit
residuals. The pulse profiles are very sinusoidal and their fractional
amplitude increases linearly with energy and no second harmonic is
detected. The reason why this pulsar is intermittent is still unknown
but we can rule out a one-to-one correspondence between Type I X-ray
bursts and the appearance of the pulsations.

\end{abstract}

\section{Introduction}

Recently it has been shown (\citealt{Kaaret}, \citealt{Galloway07},
\citealt{Casella}, \citealt{Gavriil}, \citealt{Altamirano}) that
intermittent pulsations can be observed during the transient outbursts
of some accreting millisecond pulsars (AMPs). It was emphasized
(\citealt{Casella}, \citealt{Gavriil}, \citealt{Altamirano}) that a
new class of pulsators is emerging with a variety of phenomenology
i.e., pulsations lasting only for the first two months of the outburst
(HETE J1900.1-2455, \citealt{Kaaret}, \citealt{Galloway07}), switching
on and off throughout the observations (SAX J1748.9-2021,
\citealt{Altamirano}, and HETE J1900.1-2455 \citealt{Galloway07}), or
even appearing for only ~150 s over 1.3 Ms of observations (Aql X-1,
\citealt{Casella}).  This is different from the behavior of other
known AMPs, where the pulsations persist throughout the outburst until
they drop below the sensitivity level of the instrumentation.


\citet{Galloway07} and \citet{Altamirano} pointed out the clear,
albeit non-trivial, connection between the appearance of pulsations in
HETE J1900.1-2455 and SAX J1748.9-2021 and the occurrence of
thermonuclear bursts (Type I X-ray bursts). \citet{Galloway07}
demonstrated that increases of pulse amplitude in HETE J1900.0$-$2455
sometimes but not always follow the occurrence of Type I
bursts nor do all bursts trigger an increase in pulse
amplitude. \citet{Altamirano} showed that in \saxjfull\ the pulsations
can re-appear after an absence of pulsations of a few orbital periods
or even a few hundred seconds. All the pulsing episodes occur close in
time to the detection of a Type I X-ray burst, but again not all the
Type I bursts are followed by pulsations so it is unclear if the
pulsations are indeed triggered by Type I X-ray bursts.\\ The pulse
energy dependence and the spectral state in which pulsations are
detected also differ between these objects. As shown by \citet{Cui},
\citet{Galloway05}, \citet{Gierlinski} and \citet{Galloway07}, the
fractional amplitude of the pulsations in AMPs monotonically decreases
with increasing energies between 2 and $\approx 20$keV. However in IGR
J00291+5934, \citet{Falanga} found a slight decrease in the pulse
amplitude between 5 and 8 keV followed by an increase up to energies
of 100 keV. For the single pulse episode of Aql X-1 the pulsation
amplitude increased with energy between 2 and $\approx 20$keV
(\citealt{Casella}). Interestingly in Aql X-1 the pulsations appeared
during the soft state, contrary to all the other known AMPs which
pulsate solely in their hard state (although many of these objects
have not been observed in a soft state). Therefore it is interesting
to investigate these issues for \saxjfull\ as well.\\ 
In order to study these various aspects of \saxjfull\ an
improved timing solution is necessary. The spin frequency of the
neutron star and the orbital parameters of \saxjfull\ were measured by
\citet{Altamirano} using a frequency-domain technique that measures
the Doppler shift of the pulsar spin period due to the orbital motion
of the neutron star around the center of mass of the binary. The
source was found to be spinning at a frequency of 442.361 Hz
in a binary with an orbit of ~8.7 hrs and a donor companion mass of
$\sim 0.1-1M_{\odot}$. However, the timing solution obtained with this
technique was not sufficiently precise to phase connect between epochs
of visible pulsations. The intermittency of the pulsations creates
many gaps between measurable pulse arrival times, complicating the use
of standard techniques to phase connect the times of arrival
(TOAs). The current work presents a follow-up study to the discovery
paper of intermittent pulsations in \saxjfull\ (\citealt{Altamirano})
and provides for the first time a timing solution obtained by phase
connecting the pulsations observed during the 2001 outburst. In \S 2
we present the observations and in \S 3 we explain the technique
employed to phase connect the pulsations. In \S 4 we show our results
and in \S 5 we discuss them. In \S 6 we outline our conclusions.

\section{X-ray observations and data reduction}

We reduced all the pointed observations from the {\it{RXTE}}
Proportional Counter Array (PCA, \citealt{Jahoda}) that cover the
1998, 2001 and 2005 outbursts of \saxjfull. The PCA instrument
provides an array of five proportional counter units (PCUs) with a
collecting area of 1200 $\rm\,cm^{2}$ per unit operating in the 2-60
keV range and a field of view of $\approx 1^{\circ}$. The number of
active PCUs during the observations varied between two and
five. 
We used all the data available in the Event mode with a time resolution
of $122\rm\,\mu s$ and 64 binned energy channels and in the Good Xenon
mode with a time resolution of $1\rm\,\mu s$ and 256 unbinned
channels; the latter were re-binned in time to a resolution of $
122\rm\,\mu s$. All the observations are listed in Table 1.

\begin{table}
\caption{Observations analyzed for each outburst\label{tbl:obslist}}

\scriptsize
\begin{tabular}{lrrrl}
\hline
\hline
Outburst & Start & End & Time & Observation IDs \\
 (year) & (MJD) & (MJD) & (ks) &\\
1998 & 51051.3  & 51051.6  & 14.8 & {\tt 30425-01-01-00}\\
2001 & 52138.8  & 52198.4  & 138.0 & {\tt 60035-*-*-*}~$-$~{\tt 60084-*-*-*}\\
2005 & 53523.0 & 53581.4   & 82.2 & {\tt 91050-*-*-*}\\
\hline
\end{tabular}

\label{table1:data}
\end{table}

 For obtaining the pulse timing solution we selected only photons with
energies between $\approx5$ and 24 keV to avoid the strong background at
higher energies and to avoid the region below $\approx 5$keV where the
pulsed fraction is below $\approx 1$\% rms. The use of a wider energy band 
decreases the signal to noise ratio of the pulsations.
We used only the good time
intervals excluding Earth occultations, passages of the satellite
through the South Atlantic Anomaly and intervals of unstable
pointing. We barycentered our data with the tool {\it{faxbary}} using
the JPL DE-405 ephemeris along with a spacecraft ephemeris including
fine clock corrections that together provide an absolute timing
accuracy of $\approx 3.4$ $\rm\,\mu s$ (\citealt{Rots}). The best available
source position comes from {\it{Chandra}} observations (\citealt{Zand}
and \citealt{Pooley}, see Table 2). The background was subtracted
using the FTOOL {\it{pcabackest}}.

\section{The timing technique}

The intermittent nature of \saxjfull\ requires special care when
trying to obtain a full phase connected solution (i.e., a solution
that accounts for all the spin cycles of the pulsar between periods of
visible pulsations). With pulsations detected in only $\approx 12\%$
of the on-source data of the 2001 outburst, and only a few hundred
seconds of observed pulsations during the 2005 outburst, the solution
of \citet{Altamirano} does not have the precision required to directly
phase connect the pulses. Their spin frequency uncertainty of
$\sigma_{s}=10^{-3}$Hz implies a phase uncertainty of half a spin
cycle after only $500\rm\,s$, while the gaps between pulse episodes in
the 2001 outburst were as long as $2$ $\rm\,days$.  Moreover, the low
pulsed fraction ($\simless 2.5\%$ rms in the 5-24 keV band) limits the
number of TOAs with high signal to noise ratio.\\ Since the solution
of Altamirano et al. (2008) lacks the precision required to phase
connect the TOAs, we had to find an improved initial set of ephemeris
to be use as a starting point for the phase-connection. To obtain a
better initial timing model we selected all the chunks of data where
the pulsations could clearly be detected in the average power density
spectrum (see \citealt{Altamirano} for a description of the
technique). The total amount of time intervals with visible pulsations
were M=11 with a time length $500\rm\,s\simless\rm
t_{obs,i}\rm\simless 3000\rm\,s$ in the i-th chunk, with i=1,...,M.
We call the Modified Julian Date (MJD) where the pulsations appear in
the i-th chunk as $\rm\,MJD_{start,i}$ while the final MJD where the
pulsation disappears is
$\rm\,MJD_{end,i}=\rm\,MJD_{start,i}+t_{obs,i}$. We then measured the
spin period $P_{s,i}$, the spin period derivative $\dot{P}_{s,i}$ and,
when required, the spin period second derivative $\ddot{P}_{s,i}$ to
align the phases of the pulsations (folded in a profile of 20 bins)
for each chunk of data, i.e., to keep any residual phase drift to less
than one phase bin (0.05 cycles) over the full length $t_{obs,i}$.
These three spin parameters are a mere \emph{local} measure of the
spin variation as a consequence of, primarily, the orbital Doppler
shifts and cannot be used to predict pulse phases outside the i-th
chunk of data; they provide a 'local phase-connection' of the pulses
within the i-th chunk. Using this set of $3\times\rm\,M$ parameters we
generated a series of spin periods $P_{s}\left(t\right)$ in each i-th
chunk of data with $t\in\left[\rm\,MJD_{start,i},
\rm\,MJD_{end,i}\right]$.  The predicted spin period of the pulsar
$P_{s}$ at time $t$ is given by the equation:
\begin{equation}
P_{s}\left(t\right)=P_{s,i}+\dot{P}_{s,i}t+\frac{1}{2}\ddot{P}_{s,i}t^{2}
\end{equation} 
with $i=1...M$. We then fitted a sinusoid representing a circular Keplerian
orbit to all the predicted spin periods $P_{s}\left(t\right)$
and obtained an improved set of orbital and spin parameters. 

With this new solution we folded 300$\rm\,s$ intervals of data to
create a new series of pulse profiles with higher signal to noise
ratio. We then fitted a sinusoid plus a constant to these folded
profiles and selected only those profiles with a S/N (defined as the
ratio between the pulse amplitude and its 1$\sigma$ statistical error)
larger than 3.4. With a value of $\rm\,S/N\greateq 3.4$ we expect less
than one false detection when considering the total number of folded
profiles.  The TOAs were then obtained by cross-correlating these
significant folded profiles with a pure sinusoid.  The determination
of pulse TOAs and their statistical uncertainties closely resembles
the standard radio pulsar technique (see for example
\citealt{Taylor}). \\ If the pulsar is isolated and the measured TOAs
are error-free, the pulse phase $\phi$ at time \emph{t} can be
expressed as a polynomial expansion:
\begin{equation}\label{eq:phi}
\phi_{P}\left(t\right)=\phi_{0} + \nu_{0}(t-t_{0}) + \frac{1}{2}\dot{\nu}_{0}(t-t_{0})^{2} + ...
\end{equation}

where the subscript ``0'' refers to the epoch time $t_{0}$.  In an
AMP, the quantities $\nu_{0}$ and $\dot{\nu}_{0}$ are the spin
frequency of the pulsar and the spin frequency derivative (related to
the spin torque), respectively.  We note that the detection of a spin
up/down in AMPs is still an open issue, since many physical processes
can mimic a spin up/down (see for example \citealt{Hartman}, but see
also \citealt{Burderi}, \citealt{Papitto}, \citealt{Riggio} for a
different point of view).  Indeed in the real situation the pulse
phase of an AMP can be expressed as:
\begin{equation}\label{eq:phitot}
\phi\left(t\right) = \phi_{P}\left(t\right) + \phi_{O}\left(t\right) +
\phi_{M}\left(t\right) + \phi_{N}\left(t\right)
\end{equation}
with the subscripts P, O, M, and N referring respectively to the
polynomial expansion in eq.~\ref{eq:phi} (truncated at the second
order), the orbital and the measurement-error components and any
intrinsic timing noise of unknown origin that might be included in the
data. For example the timing noise is observed in young isolated radio
pulsars (e.g., \citealt{Groth}, \citealt{Cordes}) or has been detected
as red noise in SAX J1808.4-3658 (\citealt{Hartman}) or in high mass
X-ray binaries (\citealt{Bildsten97}).  The term
$\phi_{O}\left(t\right)$ takes into account the phase variation
introduced by the orbital motion of the pulsar around the companion.
We expect that the error-measurement component
$\phi_{M}\left(t\right)$ is given by a set of independent values and
is normally distributed with an amplitude that can be predicted from
counting statistics (e.g., \citealt{Taylor}). Fitting the TOAs with a
constant frequency model and a zero-eccentricity orbit we phase
connected the 2001 outburst pulsations in a few iterations, re-folding
the profiles with the improved solution and fitting the TOAs using the
software package TEMPO2 \citep{Hobbs}.
No significant improvement is obtained fitting for a spin derivative
and an eccentricity (using the ELL1 model in TEMPO2). We did not fit
for the orbital period in TEMPO2, because our initial fit, described
earlier in this section, included data from the 2005 outburst and thus
provided more precision on this parameter. A consistency check was
made folding the pulsations of the 2005 outburst with the two
different solutions. With the phase-connected solution of the 2001
outburst we were not able to properly fold the pulsations in the 2005
outburst, due to the relatively large error in the fitted orbital
period, whereas with the hybrid solution (pulse spin frequency $\nu$,
epoch of ascending node passage $\rm\,T_{asc}$, and projected
semi-major axis $a_{x}\rm\,sin$$\,i$ from the TEMPO2 fit to 2001 only; 
$\rm\, P_{orb}$ kept from the sinusoidal fit to
$P_{s}\left(t\right)$ for all the data) we were able to fold the pulsations
and obtained a $\approx 8\rm\,\sigma$ pulse detection with approximately
500$\rm\, s$ of the 2005 outburst data.

\section{Results}

\subsection{Timing solution}

\begin{figure}
  \begin{center}
    \rotatebox{-90}{\includegraphics[width=0.7\columnwidth]{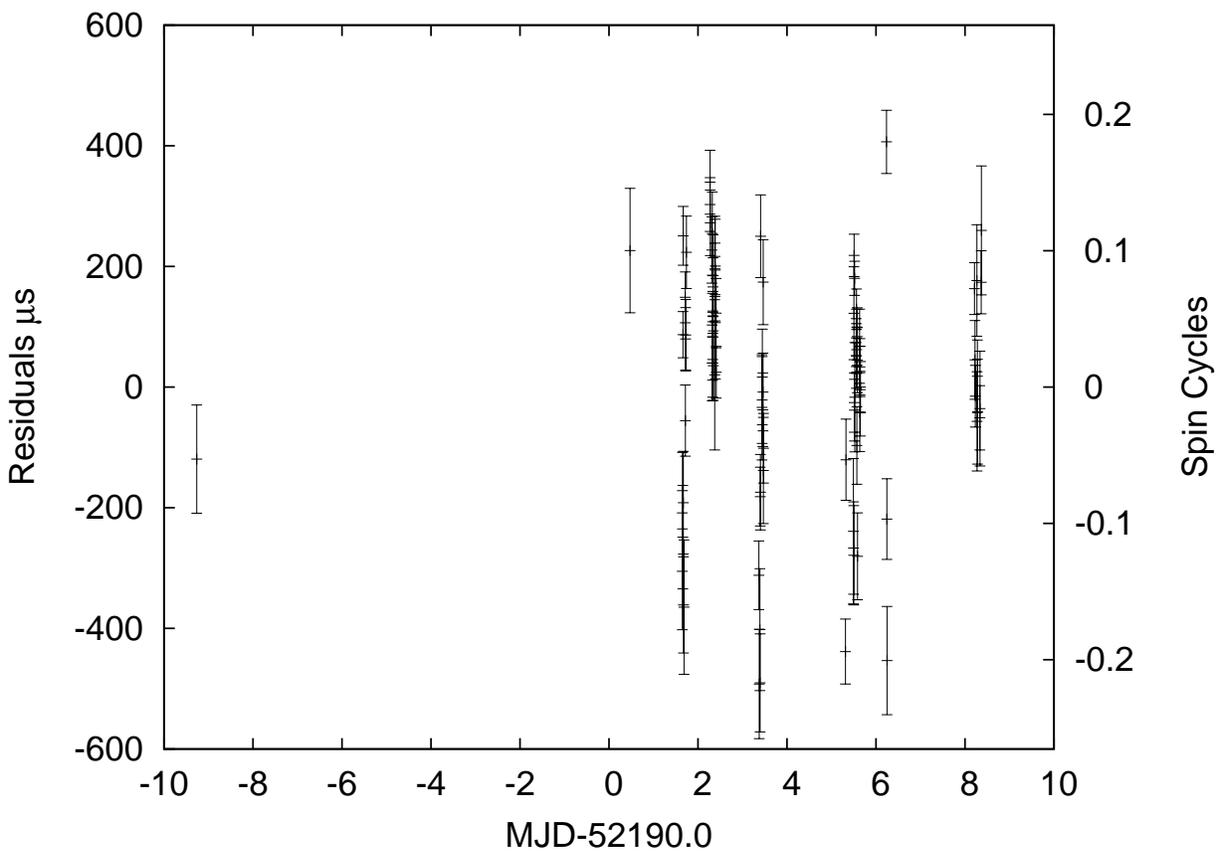}}
  \end{center}
  \caption{Post-fit timing residuals of the 5-24 keV TOAs of the 2001
outburst.  A negative/positive value of the residuals means that a
pulsation is leading/ lagging with respect to the timing model.  All
the pulse profiles used have $\rm\,S/N > 3.4$.  Strong timing noise
appears as a large scatter within each group of points and
misalignment between groups.
    \label{res}}
\end{figure}

\begin{figure}[t]
  \begin{center}
  \rotatebox{-90}{\includegraphics[width=0.7\columnwidth]{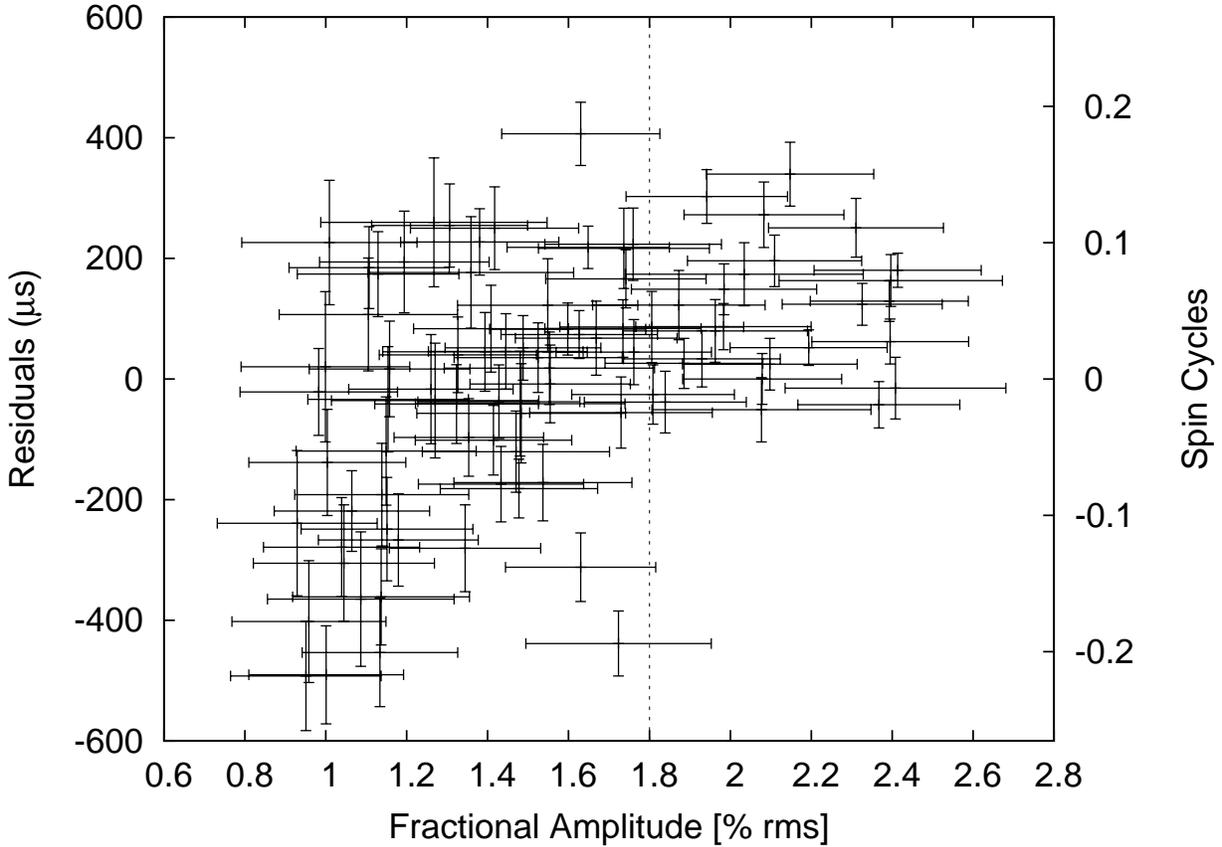}}
  \end{center}
  \caption{Dependence of the residuals on fractional amplitude of the 
pulsations. The vertical dashed line divides the plot in the upper (right side of the
panel) and
lower group (left side) of TOAs at a fractional amplitude of 1.8$\%$ rms.
\label{banana}}
\end{figure}

With the new timing solution we were able to search with higher
sensitivity for additional pulsation episodes throughout the 1998,
2001 and 2005 outbursts. We folded chunks of data with a length
between approximately $300$s (to minimize the pulse smearing due to
short timescale timing noise) and $1$ hr (to increase the S/N when the
pulsations are weak). Two new pulsation episodes of length $\approx
300$s each, were detected on MJD~52180.7 and MJD~52190.5, about eleven
days and one day before the first earlier detection respectively,
with a significance of $\approx 3.5\sigma$. We refitted a constant
frequency model and a zero eccentricity orbit to the new ensemble of
TOAs and obtained new timing residuals and a new timing solution. The
post-fit residuals are all less than $\approx 0.2$ spin cycles, but their
$\rm\, rms$ amounts to $164\rm\,\mu$s,
well in excess of the value expected from counting statistics of
$\simeq 70\mu$s; the reduced $\chi^{2}$
is $9$ (for 86 degrees of freedom, see the timing residuals in Figure~\ref{res}). Trying to fit higher frequency derivatives or an eccentricity does not
significantly improve the fit.
 At first glance, the residual variance,
described by the timing noise term $\phi_{N}$ in eq~\ref{eq:phitot},
has two components: a long-term and a short-term component to the
timing noise $\phi_{N}$. The short term timing noise acts on a
timescale of a few hundred seconds and creates the large scatter
observed within each group of points in Fig.~\ref{res}. The long term
timing noise acts on a timescale of a few days and can be recognized
by a misalignment of the TOA residuals between groups, where with a
simple white noise component we would expect within each group a
distribution of TOAs spread around zero. The short term and the long term
timing noise, which could of course well be part of the same process,
are together responsible for the bad $\chi^{2}$ of the fit and the
large $\rm\,rms$ of the residuals. \\ A histogram of the timing
residuals shows a non-Gaussian distribution with an extended lower
tail. A superposition of two Gaussians
distribution can fit this. The first Gaussian is composed of TOAs that
are on average earlier than predicted by the timing solution. The
second Gaussian comprises TOAs that are on average later than
predicted. The separation between the two mean values of the Gaussians
is a few hundred microseconds. We found that the
lagging TOAs correspond to systematically higher S/N pulsations than
the leading ones. Since pulsations with higher S/N usually have high
fractional amplitude, we analyzed the fractional amplitude dependence
of the residuals. As can be seen in Fig.~\ref{banana} above a
fractional amplitude of $\approx 1.8\%$ rms (the ``upper group''), the TOAs
are on average 103$\rm\,\mu$s later than predicted by the timing
solution; the remaining TOAs (the ``lower group'') are
42$\rm\,\mu$s early. The lower group TOAs are not
related in a one-to-one relation with the fractional amplitude
(Fig.~\ref{banana}), meaning that probably some kind of noise process
is affecting the lower group TOAs in addition to the effect the
varying amplitude has on the TOAs. 

\begin{figure*}[t]
  \begin{center}
    \rotatebox{-90}{\includegraphics[width=0.3\textwidth]{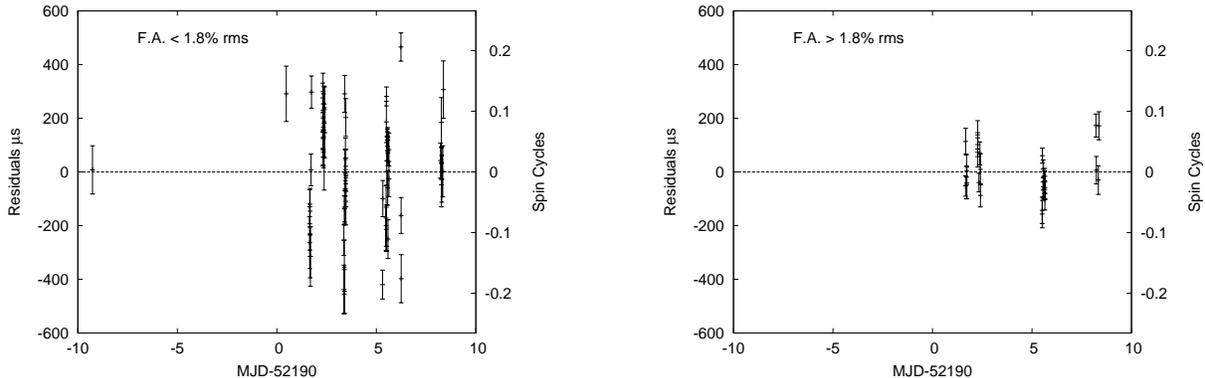}}
  \end{center}
  \caption{Timing residuals of the upper (right panel) and lower (left
panel) group. All the TOAs have at least a S/N larger than 3.4 and
refer to the 5-24 keV energy band. The short timescale timing noise
affects more strongly the lower group than the upper one, which is
reflected in the bad $\chi^{2}$ of the fit (4.4 and 9.6 for the upper
and lower group respectively). The new TOA around MJD 52181 belong to
the lower group. The two plots have the same scale on the
residuals-axis to put in evidence the large scatter of the post-fit
residuals in the lower group with respect to the upper one.
\label{res_two}}
\end{figure*}
We fitted separate constant frequency and zero eccentricity
models to just the upper and lower groups of TOAs and again checked the
post-fit residuals (see Fig.~\ref{res_two}).
The reduced $\chi^{2}$ values of the two solutions are still large
($\approx4.4$ and $\approx9.6$ for the upper and lower solution,
respectively). The maximum residual amplitude is about 200$\rm\,\mu$s
and 450$\rm\,\mu$s in the upper and lower solution, respectively.
Long term timing noise is still apparent, contributing more to the
lower TOAs group than to the upper one. While the fitted orbital
parameters ($\rm\,T_{asc}$ and $a_{x}\rm\,sin$$\,i$) are in agreement
to within 1$\sigma$, the two solutions deviate in pulse frequency by
$1.2\times10^{-7}\rm\,Hz$, about $4\sigma$. 

We ascribe this
difference to the unmodeled timing noise component $\phi_{N}$, which is
ignored in the estimate of the parameter errors. Selecting groups of
TOAs with fractional amplitude thresholds different from $1.8\%$ rms
always gives different timing solutions, with deviations larger than
$3\sigma$ sometimes also in $\rm\,T_{asc}$ and
$a_{x}\rm\,sin$$\,i$. 

To investigate if this indicates a systematic
connection between timing noise excursions and fractional amplitude or
is just a consequence of selecting subsets of data out of a record
affected by systematic noise, we selected subsets of data using
different criteria: we split the data in chunks with TOAs earlier and
later than a fixed MJD or we selected only TOAs with S/N larger than
3.4. In each case the fitted solutions deviated by several standard
deviations, indicating that the differences are due to random
selection of data segments out of a record affected by correlated
noise (the timing noise).  To take this effect into account in
reporting the timing solution we increased our statistical errors by a
constant factor such that the reduced $\chi^{2}$ of the timing
solution fit was close to unity and refitted the parameters. 
The parameters and the errors for the global solution, calculated in
this way, are reported in Table 2. They provide an improvement of
three and five orders of magnitude with respect to the Keplerian and
spin frequency parameter uncertainties, respectively, when compared
with the solution reported in \citet{Altamirano}.

\begin{table}
\caption{Timing parameters for \saxjfull\ }

\scriptsize
\begin{tabular}{lc}
\hline
\hline
Parameter & Value \\
\hline
Right Ascension ($\alpha$) (J2000)$^{a}$ & $17^{h}48^{m}52^{s}.163$ \\ 
Declination ($\delta$) (J2000)$^{a}$ & $-20^{\circ}21^{'}32^{''}.40 $\\
Orbital period, P$_{orb}$(hr) \dotfill                           & 8.76525(3) hr \\
Projected semi major axis,  $a_\mathrm{x} \sin i$ (light-ms)\dotfill   & 387.60(4) \\
Epoch of ascending node passage $T_\mathrm{asc}$ (MJD, TDB) \dotfill          & 52191.507190(4)  \\
Eccentricity, $e$ (95\% confidence upper limit) \dotfill                                         & $<2.3\times 10^{-4}$ \\
Spin frequency $\nu_0$ (Hz) \dotfill                             & 442.36108118(5) \\
Reference Epoch (MJD) & 52190.0\\
\hline
All the uncertainties quoted correspond to 1$\sigma$ confidence level (i.e., $\Delta\chi^{2}=1$).\\
$^{a}$X-ray position from Chandra \citep{Zand,Pooley}. The
pointing uncertainty is 0''.6.
\end{tabular}

\label{table2:data}
\end{table}

To check whether the timing noise is energy dependent, we selected three
energy bands: a soft band (5-13 keV), an intermediate band (11-17 keV)
and a hard one (13-24 keV). Then we repeated the procedure, fitting
all the significant TOAs with a constant spin frequency and circular
Keplerian orbit, without increasing the statistical errors on the
TOAs. 

The two solutions of the soft and intermediate bands were in
agreement to within 1$\sigma$ for $\nu$, $\rm\,T_{asc}$ and
$a_{x}\rm\,sin$$\,i$, while the hard band solution deviated by more
than 3$\sigma$ for the spin frequency. The reduced $\chi^{2}$ of the
sinusoidal fit to the pulse profiles in the highest energy band was
systematically larger than in the other two bands. Fitting a second
and a third harmonic did not significantly improve the fit according
to an F-test. 

The bad $\chi^{2}$ can be ascribed to a non-Poissonian
noise process that dominates at high energies or to an intrinsic
non-sinusoidal shape of the pulse profiles. Since the hard energy band
has a count rate approximately 10 times smaller than the soft band,
the influence of this on the global timing solution calculated for the
energy band 5-24 keV is negligible. We refolded all the profiles and
calculated another solution for the 5-17 keV energy band to avoid the
region of high energies and found an agreement to within 1$\sigma$
with the solution reported in Table 2 and with the solution obtained
for the 5-13 keV and for the 11-17 keV energy bands. This also shows
that the timing noise is substantially independent from the energy
band selected for energies below 17 keV.

We checked for the possible existence of characteristic frequencies in
a power density spectrum (PDS) of the timing residuals using a
Lomb-Scargle technique. No peaks above three sigma were found in
either the PDS of the upper and lower group TOAs or the PDS of the
global solution combining these two groups. The post-fit timing
residuals are also not correlated with the reduced $\chi^{2}$ of each
fitted pulse profile.  We also checked the influence of very large
events (VLE) due to energetic particles in the detector on the TOA
residuals but no correlation was found. There is also no link between
the source flux and the magnitude of the timing residuals and between
the timing residuals and the orbital phase. 

\subsection{Pulse shape variations}\label{sec:shape}

To look for possible pulse profile changes, we fitted all significant
folded pulse profiles with a fundamental and a second harmonic
sinusoid (plus a constant level). No significant second harmonic
detections ($>3\rm\sigma$) were made in the 5-24 keV energy band with
upper limits of 0.9\%, 0.5\% and 0.4\% rms amplitude at 98$\%$
confidence for the 1998, 2001 and 2005 pulsations episodes
respectively. We repeated the procedure for different energy bands,
obtaining similar results.\\ The pulse profiles of \saxjfull\ are
therefore extremely sinusoidal with the fundamental amplitude at least
$\approx 11$ times that of the second harmonic 
both in the soft and hard energy bands. Thus there are no detectable pulse
profile shape variations nor the possibility to detect sudden changes
of the relative phase between the fundamental and the second harmonic
such as observed in SAX J1808.4-3658 (\citealt{Burderi},
\citealt{Hartman}) and XTE J1814-338 \citep{Papitto}.

\subsection{Pulse energy dependence and time lags}

\begin{figure}[t]
  \begin{center}
  \rotatebox{-90}{\includegraphics[width=0.3\columnwidth]{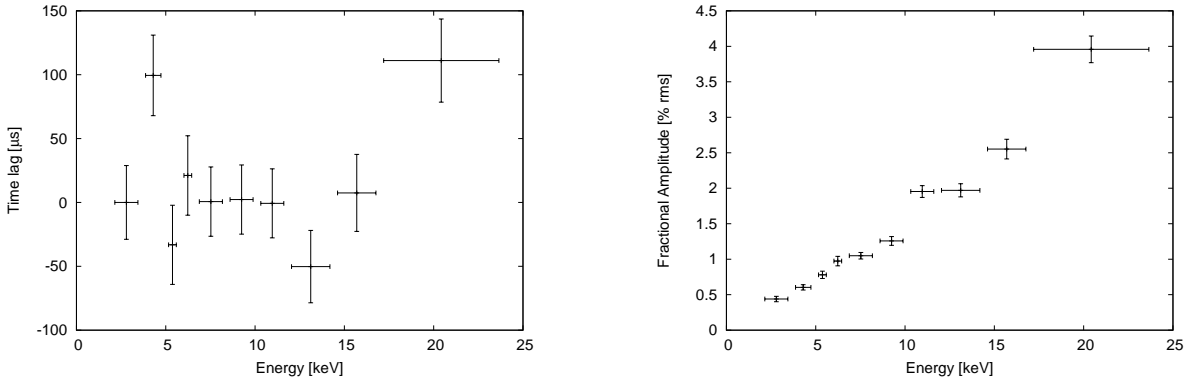}}
  \end{center}
  \caption{Time lags (left panel) and energy dependence of the
fractional amplitudes of the pulsations (right panel).  The plots stop
at 24 keV after which the counts strongly drop below the detection
level. The plots refer to all the observations where we detected
pulsations.  \textbf{Left}: no significant time lags are measured,
with all the points being consistent with a zero lag at the three
sigma level with respect to the first energy band, chosen here to be
our reference time. \textbf{Right}: the fractional amplitude starts
with a very low value of $\approx 0.5\%$ rms around 2 keV and linearly
increases up to 4$\%$ rms at around 15 keV. 
\label{energy}}
\end{figure}

We analyzed the energy dependence of the pulse profiles selecting ten
energy bands and measuring the strength of the pulsations. We folded
all the observations where we detect significant pulsations. The pulse
fraction is as small as $\approx 0.5\%$ rms between 2.5 and 4 keV and
increases up to $\approx4.0\%$ rms above 17 keV following a linear trend.
Counting statistics prevents the measurement of pulsations above 24
keV. Fitting a linear relation to the points in
Fig.~\ref{energy} gives a slope of $(0.17\pm 0.01)\rm\,
\%\,rms\,keV^{-1}$. 

The pulses of \saxjfull\ exhibit a similar energy dependence to
Aql X-1 \citep{Casella}, which is opposite to that measured in three
other AMPs (SAX J1808.4-3658 \citealt{Cui}, XTE J1751-305
\citealt{Gierlinski}, XTE J0929-314 \citealt{Galloway07}), where the
amplitude is stronger at low energies and decreases at high
energies. In another source however (IGR J00291+5934) \citet{Falanga}
pointed out an increase of the fractional amplitude with energy which
seems to resemble what has been observed for \saxjfull\ and Aql
X-1. In IGR J00291+5934 a slight decrease in the fractional amplitude
between 5 and 8 keV is followed by a slight increase up to 100
keV. However between 6 and 24 keV the slight increase is also
consistent with a constant when considering the 1$\sigma$ error bars.
In \saxjfull\ and Aql X-1 the increase in fractional amplitude with
energy is much steeper. 

We did not detect any significant time lag between soft and hard
photons, with an upper limit of $250\mu$s at the $3\sigma$ level,
using a coherent analysis between the 2.5 and 24 keV bands
(see fig.~\ref{energy}).
If a time lag with a magnitude smaller than our limit exists, this
might explain the dependence of TOA on pulse amplitude if for example
low amplitude pulses are dominated by soft photons whereas the high
amplitude pulses are dominated by hard photons or vice versa. To check
this possibility we re-folded all the low and high amplitude pulses
selecting a low energy (5-13 keV) and an high energy band (13-24
keV). The folded profiles do not show any strong energy dependence,
with the low and high energy bands equally contributing to the low and
high amplitude profiles. Therefore energy dependent lags are unlikely
to be the origin of the TOA dependence on amplitude.

\subsection{Pulse dependence on flux}
\begin{figure}[t!]
  \begin{center}
  \rotatebox{-90}{\includegraphics[width=0.7\columnwidth]{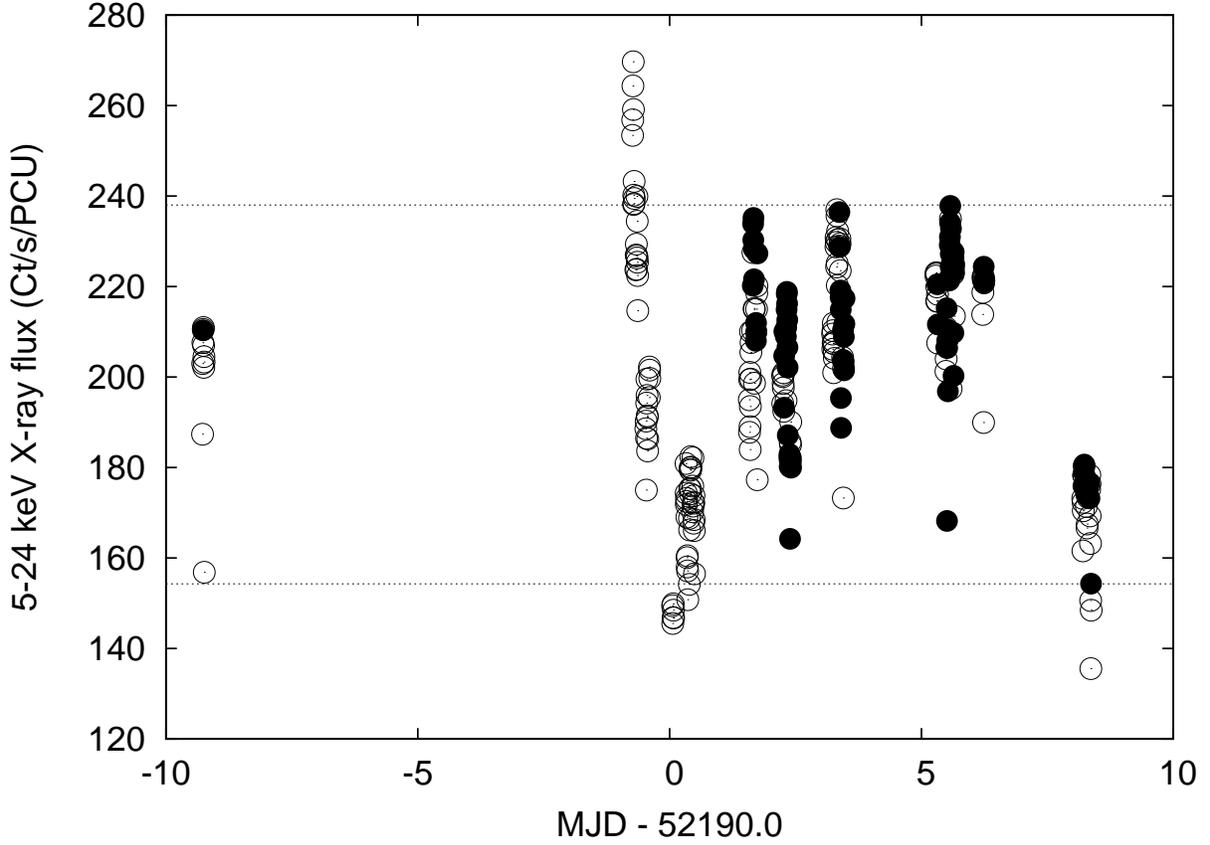}}
  \end{center}
  \caption{Light curve of the 2001 outburst from the first pulsating
episode until the end of the outburst. The time resolution of the
light curve is 300 s, and each point corresponds to a folded light
curve. When a pulsation is detected a filled circle is plotted. The
dotted lines are limits of highest and lowest flux where the pulses
appear. Note that even between the dashed lines, there are
non-pulsating episodes. These are often followed by pulsations on a
timescale of a few hundred seconds.
    \label{lc_pulses}}
\end{figure}

In both the 2001 and 2005 outbursts the pulsations disappear at both
low and high X-ray flux. In Fig.~\ref{lc_pulses} we show the light
curve of the 2001 outburst, with pulsating episodes indicated with
filled circles. The pulses appear in a broad flux band but not below
$\approx 150$ and not above $\approx 240\rm\,Ct/s/PCU$. It is also
intriguing that pulsations sporadically disappear on a timescale of a
few hundred seconds even when the flux is in the band where we see
pulsations. The few non-pulsating episodes at high and low flux could
be random, however some non-pulsing episodes are observed at the same
fluxes where we observe pulsations at different times.  This indicates
the absence of a flux threshold above or below which the pulse
formation mechanism is at work.

\subsection{Type I X-ray bursts}

\begin{figure*}
  \begin{center}
  \rotatebox{0}{\includegraphics[width=1.0\textwidth]{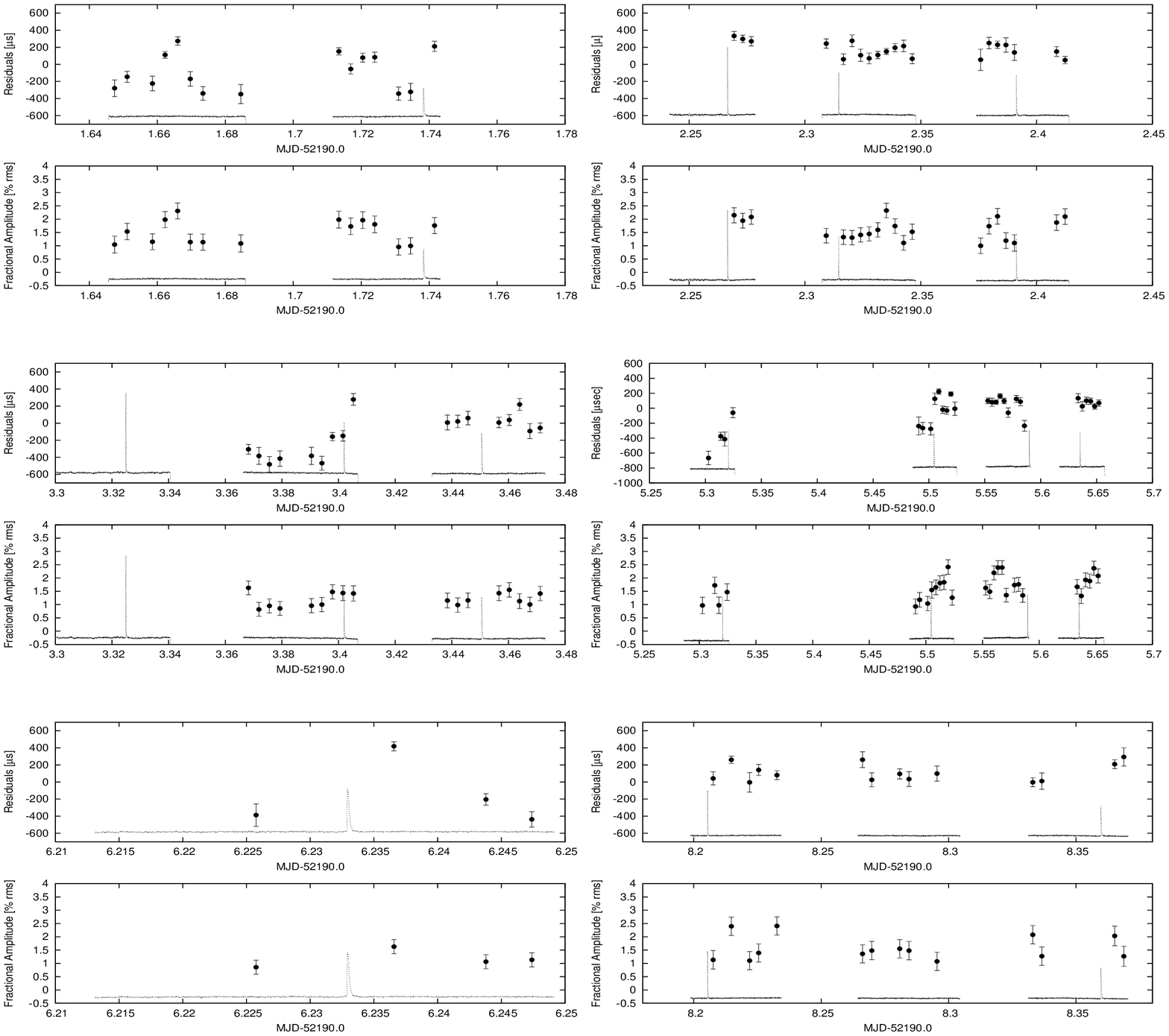}}
  \end{center}
  \caption{Time dependence of the fractional amplitude and the TOA
residuals in the 5-24 keV band during the 2001 outburst. Each diagram
displays the residuals (upper panel)  and their fractional amplitude (lower panel)
versus time. All the points correspond to pulses with a
$\rm\,S/N\greateq 3.4$. Type I bursts are also plotted in both
panels in arbitrary units but scaled by a common constant factor.
    \label{resamp}}
\end{figure*}

The possibility that the pulsations are triggered by Type I X-ray
bursts can be tested by measuring the evolution of the fractional
amplitude of the pulsations in time. Our time resolution corresponds
to the time interval selected to fold the data, i.e., $\approx300$s. As
can be seen in Fig.~\ref{resamp}, there are cases in which the
fractional amplitude increases after a Type I burst (e.g., MJD
52191.74, 52192.25, 52192.31, 52196.24, 52198.2, 52198.36, 53534.46)
on timescales of a few hundred seconds, while after other Type I
bursts the fractional amplitude does not change significantly (MJD
52190.38, 52190.47, 52193.32, 52193.40, 52193.45, 52195.3, 52195.5) or
increased prior to the Type I burst (MJD 52195.62). In four cases
pulsations are present without any occurrence of a Type I burst (MJD
52181,52190.30, 52191.65, 52198.27) and in one case the fractional
amplitude decreases after the burst (MJD 52192.40). Clearly there is
no single response of the pulsation characteristics to the occurrence
of a burst. There is also no clear link between the strength of a Type
I burst and the increase of the pulsation amplitudes as is evident
from Fig.~\ref{resamp}. The maximum fractional amplitude ($3.5\%$ rms)
is reached during the 2005 outburst, after a Type I X-ray
burst. However, after $\approx 500$s the pulsations were not seen
anymore.

\subsection{Color--color diagram}\label{sec:ccd}

We use the 16 s time resolution Standard 2 mode data to calculate
X-ray colors. Hard and soft colors are defined respectively as the
9.7--16.0 keV / 6.0--9.7 keV and the 3.5--6.0 keV / 2.0--3.5 keV count
rate ratio. The energy-channel conversion is done using the
pca\_e2c\_e05v02 table provided by the RXTE Team.
Type I X-ray bursts were removed, background
subtracted, and dead-time corrections made. In order to correct for
the gain changes as well as the differences in effective area between
the PCUs themselves, we normalized our colors by the corresponding
Crab Nebula color values \cite[see][ see table 2 in
\citealt{Altamirano} for average colors of the Crab Nebula per
PCU]{Kuulkers,Straaten} that are closest in time but in the same
RXTE gain epoch, i.e., with the same high voltage setting of the PCUs
\citet{Jahoda}.

\begin{figure}
 \begin{center}
   \rotatebox{-90}{\includegraphics[width=0.7\columnwidth]{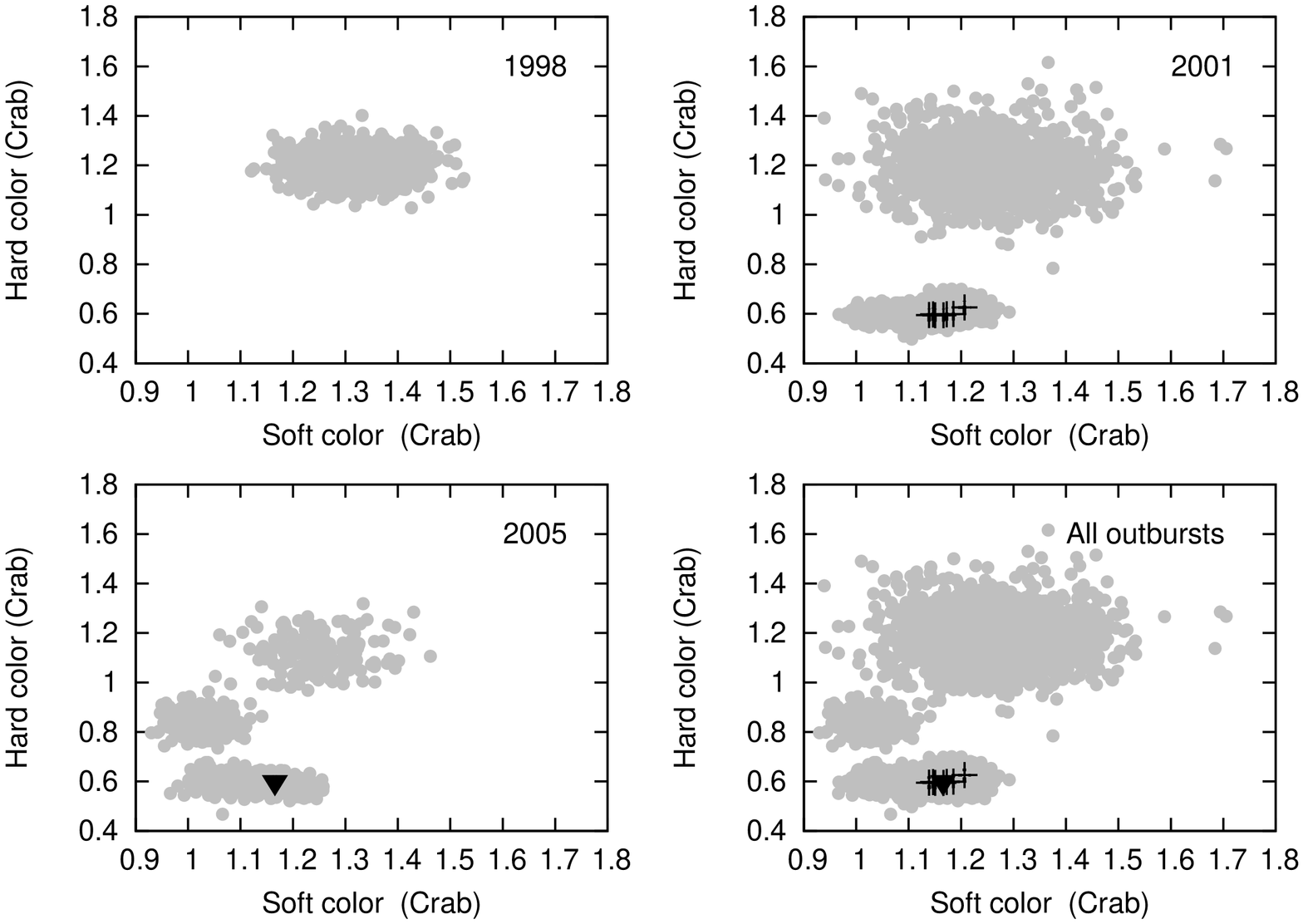}}
 \end{center}
 \caption{Color-color diagram for the 1998, 2001 and 2005 outbursts
 (values are normalized by the Crab Nebula, see
 Section~\ref{sec:ccd}). Gray circles mark the 16 seconds average
 color of all available data. Black crosses (2001 outburst) and
 triangles (2005 outburst) mark the average color per observation in
 which we find pulsations.
   \label{ccd}}
\end{figure}

The PCA observations sample the source behavior during three different
outbursts \citep[see Fig 1 in][]{Altamirano}. In Figure~\ref{ccd} we
show the color--color diagram for all the observations of the three
outbursts.
Grey circles mark the 16 second averaged colors while black crosses
and triangles mark the average color of each of the 8 observations
from which pulsations were detected. As it can be seen, pulsations
only appear in soft state of the source (banana state). During
the observations of the 1998 outburst, the source was always in the so called
Island/Extreme island state while during the 2001 and 2005 outburst the
source was observed in both island and banana state.


\subsection{Radio pulse search}
We also used the new phase-coherent timing solution presented here to search
for possible radio pulsations from SAX J1748.9$-$2021. For this, we used
2-GHz archival radio data from Green Bank Telescope (GBT) observations of
the 6 known radio millisecond pulsars (MSPs) in the globular cluster (GC)
NGC~6440 (see \citealt{Freire} for a description of these data). The
known radio pulsars in NGC~6440 have spin periods from 3.8$-$288.6\,ms and
dispersion measures (DMs) between 219$-$227\,pc cm$^{-3}$. With a spin
period of 2.3\,ms, SAX~J1748.9$-$2021 is thus the fastest-spinning pulsar
known in NGC~6440, where the average spin period of the radio pulsars
(11.3\,ms when the 288-ms pulsar B1745$-$20 is excluded) is relatively long
compared to other GCs.

We searched 11 data sets taken between 2006 December and 2007 March. These
data sets have total integration times between $0.5-5$\,hr and a combined
total time of 20.3\,hr (73\,ks). The data were dedispersed into 10 time
series with trial DMs between 219$-$228\,pc cm$^{-3}$. To check if
pulsations were present, these time series were then folded with the timing
solution presented in this paper. Because potential radio pulsations may
also be transient, we folded not only the full data sets, but also
overlapping chunks of 1/4, 1/10, and 1/50 of the individual data set (thus
probing timescales between 100\,s and hours). Furthermore, because these data
were taken 6 years after the data that was used to construct the timing
solution, we folded the data using both the exact period prediction from the
timing solution as well as allowing for a small search around this value.

No obvious pulsations were detected in this analysis, where the
reduced $\chi^2$ of the integrated pulse profile was used to judge if
a given fold was worthy of further investigation. This is perhaps not
surprising as radio pulsations have, as of yet, never been detected
from an LMXB or AMP. Nonetheless, we plan to continue these searches
on the large amount of available radio data which we have not yet
searched with this technique.

\section{Discussion}

We have presented a new timing solution for \saxjfull\, obtained by
phase connecting the 2001 outburst pulsations. We discovered the
presence of timing noise on short (hundred seconds) and long (few
days) timescales.  We cannot exclude the presence of timing noise on
different timescales, since the short and the long timescales found
are also the two timescales the TOAs probe. The pulse profiles of
\saxjfull\ keep their sinusoidal shape below 17 keV throughout the
outburst, but do experience considerable shifts relative to the
co-rotating reference frame both apparently randomly as a function of
time and systematically with amplitude. Above 17 keV the pulse
profiles show deviations from a sinusoidal shape that cannot be
modeled adding a 2nd harmonic. The fit needs a very high number of
harmonics to satisfactory account the shape of the pulsation.
This is probably due to the effect of some underlying unknown
non-Poissonian noise process that produces several sharp spikes in the 
pulse profiles. The lack of a detectable second harmonic
prevents us from studying shape variations of the pulse profiles such
as was done for SAX J1808.4--3658 \citep{Burderi,Hartman} where sudden
changes between the phase of the fundamental and the second harmonic
were clearly linked with the outburst phase.\\ Another interesting
aspect is the energy dependence of the pulse profiles, which can be a
test of current pulse formation theories. In one model of
AMPs, \citet{Poutanen} explain the pulsations by a modulation of
Comptonized radiation whose seeds photons come from blackbody
radiation. The thermal emission is given by the hot spot and/or
emitting column produced by the in-falling material that follows the
magnetic dipole field lines of the neutron star plasma rotating with
the surface of the neutron star. Part of the blackbody photons can be
scattered to higher energies by a slab of shocked plasma that forms a
comptonizing region above the hot spot \citep{Basko}.  

In three AMPs
\citep{Cui,Gierlinski,Poutanen,Watts06,Galloway07} the fractional
pulse amplitudes decreases toward high energies with the soft photons
always lagging the hard ones.  However in \saxjfull\ the energy
dependence is opposite, with the fractional pulse amplitudes
increasing toward higher energies and no detectable lags (although the
large upper limit of $250\rm\,\mu$s does not rule out the presence of
time lags with magnitude similar to those detected in other
AMPs). Moreover, all the pulsating episodes happen when the source is
in the soft state (although not all soft states show pulsations). This
is similar to Aql X-1 (see \citealt{Casella}) where the only pulsating
episode occurred during a soft state, and the pulsed fraction also
increased with energy. Remarkably all the other sources which show
persistent pulsations have hard colors typical of the extremely island
state. 

Both the energy dependence and the presence of the pulsations
during the soft state strongly suggest a pulse formation pattern for
these two intermittent sources which is different from that of the
other known AMPs and the intermittent pulsar HETE J1900.1-2455.  As
discussed in \citealt{MunoA}, \citealt{MunoB}, a hot spot region
emitting as a blackbody with a temperature constrast with respect to
the neutron star surface produces pulsations with an increasing
fractional amplitude with energy in the observer rest frame. The exact
variation of the fractional amplitude with energy however has a
complex dependence on several free parameters as the mass and radius
of the neutron star and the number,size, position and temperature of
the hot spot and viewing angle of the observer. The observed slope of
0.2$\rm\,\%\,rms\,keV^{-1}$ is consistent with this scenario, i.e., a
pure blackbody emission from a hot spot with a temperature constrast
and a weak comptonization (see \citealt{FalangaTit}, \citealt{MunoA},
\citealt{MunoB}). However it is not possible to exclude a
strong comptonization given the unknown initial slope of the
fractional amplitude.

The pulse shapes above 17 keV are
non-sinusoidal and are apparently affected by some non-Poissonian
noise process or can be partially produced by an emission mechanism
different from the one responsible of the formation of the soft pulses
(see for example \citet{Poutanen} for an explaination of how the soft
pulses can form).  We found that the TOAs of the pulsations are
independent of the energy band below 17 keV but selection on the
fractional amplitude of the pulsations strongly affects the TOAs: high
amplitude pulses arrive later. However, this does not affect the
timing solution beyond what is expected from fitting other data
selections: apparently the TOAs are affected by correlated timing
noise. If we take into account the timing noise ($\phi_{N}$) in
estimating the parameter errors, then the timing solutions are
consistent to within 2$\sigma$. In the following we examine several
possibilities for the physical process producing the $\phi_{N}$ term.


\subsection{Influence of Type I X-ray bursts on the TOAs}

The occurrence of Type I X-ray bursts in coincidence with the
appearance of the pulsations suggests a possible intriguing relation
between the two phenomena. However as we have seen in \S 4.4, there is
not a strict link between the appearance of pulsations and Type I
burst episodes. Only after $\approx 30\%$ of the Type I
bursts the pulsations appeared or increased their fractional
amplitude. The appearance of pulsations seems more related with a
period of global surface activity during which both the pulsations and
the Type I bursts occur. This is different from what has been observed
in HETE J1900.1-2455 \citep{Galloway07} where the Type I X-ray bursts
were followed by an increase of the pulse amplitudes that were
exponentially decaying with time. We note that during the pulse
episodes at MJD 52191.7, 52193.40, 52193.45 and 52195.5 the TOA
residuals just before and after a Type I burst are shifted by
300-500$\rm\,\mu$s suggesting a possible good candidate for the large
timing noise observed. However in two other episodes no substantial shift is
observed in the timing residuals (see Fig.~\ref{resamp}). We also note that
 there is no relation between the Type I burst peak flux
and the magnitude of the shifts in the residuals.

\subsection{Other possibilities}

The timing noise and its larger amplitude in the weak pulses might be
related with some noise process that becomes effective only when the
pulsations have a low fractional amplitude.  \citet{Romanova} have
shown that above certain critical mass transfer rates an unstable
regime of accretion can set in, giving rise to low-$m$ modes in the
accreting plasma which produce an irregular light curve. The
appearance of these modes inhibits magnetic channeling and hence
dilutes the coherent variability of the pulsations. Such a model can
explain pulse intermittency by positing that we see the pulses only
when the accretion flow is stable. The predominance of timing noise in
the low fractional amplitude group could be explained if with the
onset of unstable accretion the $m$-modes gradually set in and affect
the pulsations, lowering their fractional amplitude until they are
undetectable. However the lack of a correlation between pulse
fractional amplitude and X-ray flux is not predicted by the model.

Another interesting aspect are the apparent large jumps that we observe
in the phase residuals on a timescale of a few hundred seconds.
Similar jumps have been observed for example in SAX J1808.4-3658
during the 2002 and 2005 outbursts (\citealt{Burderi,Hartman}.  In
that case however the phases jump by approximately 0.2 spin cycles on
a timescale of a few days, while here we observe phase jumps of about
0.4 cycles on a timescale of hundred seconds.  Other AMPs where phase
jumps were observed (XTE J1814-338 \citealt{Papitto}, XTE J1807-294
\citealt{Riggio}) have shown timescales of a few days similar to SAX
J1808.4-3658. This is a further clue that the kind of noise we are
observing in \saxjfull\ is somewhat different from what has been
previously observed.

\section{Conclusions}

We have shown that it is possible to phase connect the intermittent
pulsations seen in \saxjfull\ and we have found a coherent timing
solution for the spin period of the neutron star and for the Keplerian
orbital parameters of the binary.\\ We found strong correlated timing
noise in the post-fit residuals and we discovered that this noise is
strongest in low fractional amplitude pulses and is not related with
the orbital phase. Higher-amplitude ($\simgreat 1.8\%$ rms) pulsations
arrive systematically later than lower-amplitude ($\simless 1.8\%$ rms)
ones, by on average 145$\mu$s. The pulsations of \saxjfull\ are
sinusoidal in the 5-17 keV band, with a fractional amplitude
linearly increasing in the energy range considered. The pulsations
appear when the source is in the soft state, similarly to what has
been previously found in the intermittent pulsar Aql X-1. The origin
of the intermittency is still unknown, but we can rule out a
one-to-one correspondence between Type I X-ray bursts and the
appearance of pulsations.

\acknowledgements{Acknowledgements: We would like to thank P. Soleri,
 B. Stappers and J. Verbiest for useful discussions. J.W.T.H. thanks
 NSERC and the Canadian Space Agency for a postdoctoral fellowship and
 supplement respectively.The National Radio Astronomy Observatory is a
 facility of the National Science Foundation operated under
 cooperative agreement by Associated Universities, Inc.}

\end{document}